\documentclass[sigconf]{acmart}

\AtBeginDocument{
	\providecommand\BibTeX{{
			\normalfont B\kern-0.5em{\scshape i\kern-0.25em b}\kern-0.8em\TeX}}}

\usepackage{verbatim}
\usepackage{amsfonts}
\usepackage{amsmath}
\usepackage{array}
\usepackage{subcaption}
\usepackage{balance}
\usepackage{multicol}
\usepackage{multirow}
\usepackage{color}
\usepackage{epsfig}
\usepackage{natbib}
\usepackage{xspace}
\usepackage{booktabs}
\usepackage{bm}
\usepackage{enumitem}
\usepackage{diagbox}
\usepackage{url}
\usepackage{geometry}
\usepackage{changepage}
\usepackage{textcomp}
\usepackage{graphicx}
\graphicspath{ {./pics/} }
\usepackage{color}
\usepackage{bm}
  
\usepackage{amssymb}
\usepackage{algorithm}
\usepackage{algorithmic}
\usepackage{threeparttable}
\usepackage{multirow}
\usepackage{mathrsfs}
\DeclareMathAlphabet{\mathcal}{OMS}{cmsy}{m}{n}

\newcommand{\hide}[1]{} 
\newcommand{\vpara}[1]{\noindent\textbf{#1 }}

\hypersetup{draft}

\begin{document}             
\title{MMGA: Multimodal Learning with Graph Alignment}

\author{Xuan Yang}
\affiliation{%
  \institution{Zhejiang University}
  \country{}
}
\email{xuany@zju.edu.cn}

\author{Quanjin Tao}
\affiliation{%
  \institution{Zhejiang University}
  \country{}
}
\email{taoquanjin@zju.edu.cn}

\author{Xiao Feng}
\affiliation{%
  \institution{Zhejiang University}
  \country{}
}
\email{3200104919@zju.edu.cn}

\author{Donghong Cai}
\affiliation{%
  \institution{Zhejiang University}
  \country{}
}
\email{donghongcai@zju.edu.cn}

\author{Xiang Ren}
\affiliation{%
  \institution{University of Southern California}
  \country{}
}
\email{xiangren@usc.edu}

\author{Yang Yang}
\affiliation{%
  \institution{Zhejiang University}
  \country{}
}
\email{yangya@zju.edu.cn}
\authornote{Corresponding author.}

\def\model#1{MMGA}
\def\gnnm#1{Graph Neural Network Module}
\def\rnnm#1{Recurrent Neural Network Module}



\begin{abstract}
Multimodal pre-training breaks down the modal barriers and allows the individual modalities to be mutually augmented with information, resulting in significant advances in representation learning. However, graph modality, as a very general and important form of data, cannot be easily interacted with other modalities because of its non-regular nature. In this paper, we propose MMGA (Multimodal learning with Graph Alignment), a novel multimodal pre-training framework to incorporate information from graph (social network), image and text modalities on social media to enhance user representation learning. In MMGA, a multi-step graph alignment mechanism is proposed to add the self-supervision from graph modality to optimize the image and text encoders, while using the information from the image and text modalities to guide the graph encoder learning. We conduct experiments on the dataset crawled from Instagram. The experimental results show that MMGA works well on the dataset and improves the fans prediction task's performance. We release our dataset, the first social media multimodal dataset with graph, of 60,000 users labeled with specific topics based on 2 million posts to facilitate future research.
\end{abstract}


\maketitle

\section{Introduction}

Multimodal learning has gained increasing attention in recent years as the heterogeneous data become ubiquitous in the real world.  Nowadays, many multimodal works study the question of how to utilize the relationships between the multi modalities and enhance the cross-modal representation learning, such as CLIP~\cite{DBLP:journals/corr/abs-2103-00020}, BLIP~\cite{li2022blip} and FLAVA~\cite{singh2022flava}. However, these works mostly focus on the vision-language modalities, few works pay attention to the graph modality. The graph modality is commonly exists in the real world and is very important in many applications. For example, on the social media, besides the texts and the images that posted by users, there is abundant information in graph data (such as interaction graph and following graph), which can help us better capture the interest of user and then improve the user recommendation or advertising strategy. However, it is difficult to deeply incorporate the graph modality with image and text modalities: compared with the image and text data, the graph is naturally non-euclidean structure data, which makes the existing modality fusion method unworkable. Thus, rather than the natural alignment learning method that commonly used in image-text modalities, how to incorporate the graph data into the multimodal learning is a key challenge. Though there are works incorporate the graph data into the multimodal learning, they mainly treat the graph data as an individual modality, which only use the graph structure for information aggregation after the multi-modal (text-image) fusion process~\cite{liu2021pre}\cite{wei2019mmgcn} and ignoring the cross-modality information between graph and other. 

To address the above problem, we propose the Multimodal
learning with Graph Alignment (MMGA) framework to better leverage the information in multi modalities and enlarge their mutual information. Specifically, we conduct our study on the social media dataset. On social media, there are user social networks which contains rich user social information and user posts which simultaneously contain image and text. We propose to use the graph structure in the user social network to supervise the learnt space of image and text modality. Meanwhile, we utilize the image and text representations of users to infer the relationships between users, such as the closeness between users, and then use it to support the graph learning process. In other word, we align the explicit graph structure with the implicit image semantic structure and text semantic structure respectively. In this way, we align the graph modality's space with the semantic spaces of image and text modalities; and then, we use the distance between two nodes in image and text spaces to adjust the graph weight. In this way, we use graph modality's information to add an additional supervision on text and image modality representation learning and use the text and image modalities' information to infer the user social network information (graph information). In such a multi-step progress, we fuse the graph, image and text modalities and enhance their representations with the help from other modalities. After the alignment learning, we concatenate the representations from three modalities and get better user representations.


The main contributions of our paper can be summarized as follows:

1)To our knowledge, We are the first to introduce the graph modal into the multimodal fusion area and propose a graph alignment task to fuse the multi modalities. 

2)We design an efficient (the first) graph multimodal pretrain framework, which could improve the user representation learning on social media. 

3) We create the first social Media multimodal dataset with graph, including more than 2 million posts, a million-scale graph and user/post labels.

\section{Method} 
\label{sec:model}

Here we introduce our proposed MMGA, a unified multimodal pre-training framework to learn user representation from graph, image and text modalities on social network. This section first states the problem to describe the objectives of MMGA. We then introduces our new model architecture, the multi-step graph alignment task, and other pre-training tasks.
\subsection{Problem Statement}
Here we formally define the user representation learning problem on social media. Given the social media posts (P = ${p_1,p_2,...,p_n}$), social network ($G = (V,E)$) and personal information ($X$) of a user $U$, we aim to learn the general representation $R$ of the user that could be used to conduct multiple downstream tasks, such as user content classification, post recommendation and fans prediction.
More specifically, each social media post $p_i$ multimodally contains an image $p_i^I$ and a piece of texts $p_i^T$. And the social network $G = (V,E)$ contains the user's social relationship information, where $V$ denotes the set of nodes (i.e. users), and $E$ denotes the set of edges between them (i.e. mutual following relationships). The user's personal information $X$ contains the user's statistic information, such as received likes and posting behaviour. 

\subsection{Model Architecture}
We utilize three encoders to separately encode the information from the graph, image and text modalities. And then we concatenate the multi-modal embeddings from the three encoders to get the final representation. 

\vpara{Unimodal encoder}. For image encoder, we employ a visual transformer~\cite{dosovitskiy2020image}, which divides an input image into patches and encodes them as a sequence of embeddings, with an additional [CLS] token to represent the global image feature.~\cite{li2022blip}
\begin{equation}
    R^I = Vit(P^I)
\end{equation}
For text encoder, we employ a BERT~\cite{devlin2018bert}, where a [CLS] token is appended to the beginning of the text input to summarize the sentence.~\cite{li2022blip}
\begin{equation}
    R^T = Bert(P^T)
\end{equation}

\vpara{Graph encoder}. In order to incorporate the multimodal information from image and text into the graph learning, we propose a multimodal-mixture graph encoder based on Graph Neural Network (GNN) to learn the graph representation.
In the graph, each node $u$'s feature is set as its statistical feature $x_u$, in other words, $G = (X,E)$.
\begin{equation}
    R^G = MMGNN(X,E,R^I,R^T)
\end{equation}
We will further introduce the designed graph encoder in the next section.

\subsection{Multi-step Graph Alignment Pre-training Strategy}
We propose to break down the barries between graph modality and other modalities, use each other's information to refine the multi-modal representation spaces.
Our proposed multi-step graph alignment pre-training  strategy consists of two parts, the first part is a designed multi-modal mixture graph encoder, which inject the information from image and text modalities into the graph modality. It utilize the image and text spaces' user representations to advice the graph propagation process in the graph representation learning, and thus refine the graph representation space. The second part is a graph contrastive loss on text and image encoders, which aims to utilize the graph modality's information to enhance the representation space learning of text and image modalities.

\vpara{Multimodal-mixture graph encoder}. In the propagation process of GNN on graph, the propagation weights should naturally be different: a node will be influenced to various extent by different interaction nodes around it. The traditional way to calculate the influence weights is to look into the node graph embeddings in the last GNN layer. However, as the information from image and text actually provide us with more information than only using the node features, we could use the information from the two modalities to help infer which relationship between two nodes is of more importance to our target nodes (moreover, the type of the relationship). 
Formally, for a node $u$, given the node feature $X_u$, the interaction graph $G_u={X_u,E_u}$, the current image embedding of the node $R^I_u$ and the current text embedding of the node $R^T_u$, in the propagation process, we first calculate the influence weight of an interaction edge $E_{u,v}$:
\begin{equation}
    a_{u,v} = M-Gate(X_u,X_v,R^I_u,R^T_u,R^I_v,R^T_v)
\end{equation}
\begin{equation}
    M-Gate(X_u,X_v,R^I_u,R^T_u,R^I_v,R^T_v) = W (X_u||X_v||R^I_u||R^T_u||R^I_v||R^T_v)
\end{equation}
Then in each GNN layer, we weightily propagate the information of neighbor nodes to the target node to refine its representation:
\begin{equation}
    h_{u} = h_{u}^{t, k - 1} + \sum_{v \in N(u)}a_{u,v} \cdot h_{v}
\end{equation}

\vpara{Graph contrastive loss on image and text}. After we get the image embedding and the text embedding, we aim to use the explicit graph structure to supervise the implicit representation (semantic) space structure of image and text structure.  This idea is driven from the social proof~\cite{cialdini2007influence}, that the user will inclined to have similar behavior tendency/preference, such as post preference, as his neighbors. More specifically, we use a contrastive learning based pre-training task to supervise the image and text representation spaces, aiming at making the users that are connected in the social network have more similar representations in text and image spaces than the strangers (pairs from negative sampling).
Formally, given the current image embedding of the node $R^I$ of two users $u$ and $v$, we try to minimize the graph contrastive loss:
\begin{equation}
    \hat{y}_{u,v} = Softmax(ReLU(W_2 \cdot ReLU(W_1 \cdot R^I_u||R^I_v + b_1) + b_2))
\end{equation}
\begin{equation}
    Loss_1 = -[y_ilog(\hat{y}_i)+(1-y_i)log(1-\hat{y}_i)]
\end{equation}
where $y_{u,v}$ is labelled as 1 if the user $u$ and $v$ are connected in the social network; $y_{u,v}$ is labelled as 0 if they are a negative sampling user pair.
The graph contrastive loss works in the same way on text modality.

\subsection{Pre-training Objectives}

In the pre-training stage, we design five objectives during pre-training to optimize our model parameters on the pretrained dataset.

Firstly, we jointly optimize two objectives during pre-training to train the image and text encoder, aiming to achieve the domain shifting on the new dataset from the loaded former pretrained unimodal encoders:

\vpara{Language Modeling Loss (LM)}. It aims to make the text encoder understand more information that is expressed by the text. It masks part of words (tokens) in the text and then make the text model to guess what is the token. Then it optimizes a cross entropy loss which trains the model to maximize the likelihood of the text in an autoregressive manner~\cite{li2022blip}. Following the setting of former works, we apply a label smoothing of 0.1 when computing the loss.

\vpara{Image-Text Aligning Loss (ITA)}.It aims to align the feature space of the visual transformer and the text transformer by encouraging positive image-text pairs to have similar representations in contrast
to the negative pairs.~\cite{li2022blip} It has been shown to be an effective objective for improving vision and language understanding~\cite{DBLP:journals/corr/abs-2103-00020}\cite{li2021align}.

Besides the two objectives, we adapt two pre-training tasks to train the GNN model in the pre-training stage:

\vpara{Node Feature Modeling Loss}. It masks node features and then it lets GNNs predict those node features based on neighboring structure~\cite{devlin2018bert}. Specifically, We randomly mask input node features. We then apply GNNs to obtain the corresponding node embeddings. Finally, a linear model is applied on top of embeddings to predict a masked node/edge feature.~\cite{hu2019strategies}\cite{hu2020gpt}

\vpara{Graph Structure Modeling Loss}. We mask part of edges in the graph and then we sample the connected node pairs and random unconnected node pairs (negative sampling). For each node pair in our sampled pairs, we let GNNs predict if there exists an edge between the node pair.~\cite{hu2019strategies}

\vpara{Graph Constractive Loss}. It has been introduced above, which is designed to refine the image and text representation space learning by adding a weak self-supervision from the graph structure.~\cite{hu2019strategies}

After the pre-training stage, we add a classifier layer after we get the final user multi-modal representation by concatenating the three representations from graph, text and image modalities. In the fine-tuning stage, we evaluate our framework's effectiveness by mainly two downstream task: user content classification and fans prediction.

\section{Conclusion}
We propose a novel multi-step alignment pretrain task, which could fuse the graph, image and text modalities, and enrich their mutual information.
We design an efficient (the first) graph multimodal pretrain framework, which could simultaneously refine the representation space of three modalities. The pretrained model is expected to be useful in many downstream task on social media, such as post recommendation and user relationship identification. Our framework can also be applied to other real world datasets such as review dataset and item sharing dataset.
Furthermore, we create the first social Media multimodal dataset with graph, including more than 2 million posts, a million-scale graph and user/post labels.

\balance
\bibliographystyle{ACM-Reference-Format}
\bibliography{reference}


\begin{thebibliography}{11}


\ifx \showCODEN    \undefined \def \showCODEN     #1{\unskip}     \fi
\ifx \showDOI      \undefined \def \showDOI       #1{#1}\fi
\ifx \showISBNx    \undefined \def \showISBNx     #1{\unskip}     \fi
\ifx \showISBNxiii \undefined \def \showISBNxiii  #1{\unskip}     \fi
\ifx \showISSN     \undefined \def \showISSN      #1{\unskip}     \fi
\ifx \showLCCN     \undefined \def \showLCCN      #1{\unskip}     \fi
\ifx \shownote     \undefined \def \shownote      #1{#1}          \fi
\ifx \showarticletitle \undefined \def \showarticletitle #1{#1}   \fi
\ifx \showURL      \undefined \def \showURL       {\relax}        \fi
\providecommand\bibfield[2]{#2}
\providecommand\bibinfo[2]{#2}
\providecommand\natexlab[1]{#1}
\providecommand\showeprint[2][]{arXiv:#2}

\bibitem[\protect\citeauthoryear{Cialdini and Cialdini}{Cialdini and
  Cialdini}{2007}]%
        {cialdini2007influence}
\bibfield{author}{\bibinfo{person}{Robert~B Cialdini} {and}
  \bibinfo{person}{Robert~B Cialdini}.} \bibinfo{year}{2007}\natexlab{}.
\newblock \bibinfo{booktitle}{\emph{Influence: The psychology of persuasion}}.
  Vol.~\bibinfo{volume}{55}.
\newblock \bibinfo{publisher}{Collins New York}.
\newblock


\bibitem[\protect\citeauthoryear{Devlin, Chang, Lee, and Toutanova}{Devlin
  et~al\mbox{.}}{2018}]%
        {devlin2018bert}
\bibfield{author}{\bibinfo{person}{Jacob Devlin}, \bibinfo{person}{Ming-Wei
  Chang}, \bibinfo{person}{Kenton Lee}, {and} \bibinfo{person}{Kristina
  Toutanova}.} \bibinfo{year}{2018}\natexlab{}.
\newblock \showarticletitle{Bert: Pre-training of deep bidirectional
  transformers for language understanding}.
\newblock \bibinfo{journal}{\emph{arXiv preprint arXiv:1810.04805}}
  (\bibinfo{year}{2018}).
\newblock


\bibitem[\protect\citeauthoryear{Dosovitskiy, Beyer, Kolesnikov, Weissenborn,
  Zhai, Unterthiner, Dehghani, Minderer, Heigold, Gelly,
  et~al\mbox{.}}{Dosovitskiy et~al\mbox{.}}{2020}]%
        {dosovitskiy2020image}
\bibfield{author}{\bibinfo{person}{Alexey Dosovitskiy}, \bibinfo{person}{Lucas
  Beyer}, \bibinfo{person}{Alexander Kolesnikov}, \bibinfo{person}{Dirk
  Weissenborn}, \bibinfo{person}{Xiaohua Zhai}, \bibinfo{person}{Thomas
  Unterthiner}, \bibinfo{person}{Mostafa Dehghani}, \bibinfo{person}{Matthias
  Minderer}, \bibinfo{person}{Georg Heigold}, \bibinfo{person}{Sylvain Gelly},
  {et~al\mbox{.}}} \bibinfo{year}{2020}\natexlab{}.
\newblock \showarticletitle{An image is worth 16x16 words: Transformers for
  image recognition at scale}.
\newblock \bibinfo{journal}{\emph{arXiv preprint arXiv:2010.11929}}
  (\bibinfo{year}{2020}).
\newblock


\bibitem[\protect\citeauthoryear{Hu, Liu, Gomes, Zitnik, Liang, Pande, and
  Leskovec}{Hu et~al\mbox{.}}{2019}]%
        {hu2019strategies}
\bibfield{author}{\bibinfo{person}{Weihua Hu}, \bibinfo{person}{Bowen Liu},
  \bibinfo{person}{Joseph Gomes}, \bibinfo{person}{Marinka Zitnik},
  \bibinfo{person}{Percy Liang}, \bibinfo{person}{Vijay Pande}, {and}
  \bibinfo{person}{Jure Leskovec}.} \bibinfo{year}{2019}\natexlab{}.
\newblock \showarticletitle{Strategies for pre-training graph neural networks}.
\newblock \bibinfo{journal}{\emph{arXiv preprint arXiv:1905.12265}}
  (\bibinfo{year}{2019}).
\newblock


\bibitem[\protect\citeauthoryear{Hu, Dong, Wang, Chang, and Sun}{Hu
  et~al\mbox{.}}{2020}]%
        {hu2020gpt}
\bibfield{author}{\bibinfo{person}{Ziniu Hu}, \bibinfo{person}{Yuxiao Dong},
  \bibinfo{person}{Kuansan Wang}, \bibinfo{person}{Kai-Wei Chang}, {and}
  \bibinfo{person}{Yizhou Sun}.} \bibinfo{year}{2020}\natexlab{}.
\newblock \showarticletitle{Gpt-gnn: Generative pre-training of graph neural
  networks}. In \bibinfo{booktitle}{\emph{Proceedings of the 26th ACM SIGKDD
  International Conference on Knowledge Discovery \& Data Mining}}.
  \bibinfo{pages}{1857--1867}.
\newblock


\bibitem[\protect\citeauthoryear{Li, Li, Xiong, and Hoi}{Li
  et~al\mbox{.}}{2022}]%
        {li2022blip}
\bibfield{author}{\bibinfo{person}{Junnan Li}, \bibinfo{person}{Dongxu Li},
  \bibinfo{person}{Caiming Xiong}, {and} \bibinfo{person}{Steven Hoi}.}
  \bibinfo{year}{2022}\natexlab{}.
\newblock \showarticletitle{Blip: Bootstrapping language-image pre-training for
  unified vision-language understanding and generation}.
\newblock \bibinfo{journal}{\emph{arXiv preprint arXiv:2201.12086}}
  (\bibinfo{year}{2022}).
\newblock


\bibitem[\protect\citeauthoryear{Li, Selvaraju, Gotmare, Joty, Xiong, and
  Hoi}{Li et~al\mbox{.}}{2021}]%
        {li2021align}
\bibfield{author}{\bibinfo{person}{Junnan Li}, \bibinfo{person}{Ramprasaath
  Selvaraju}, \bibinfo{person}{Akhilesh Gotmare}, \bibinfo{person}{Shafiq
  Joty}, \bibinfo{person}{Caiming Xiong}, {and} \bibinfo{person}{Steven
  Chu~Hong Hoi}.} \bibinfo{year}{2021}\natexlab{}.
\newblock \showarticletitle{Align before fuse: Vision and language
  representation learning with momentum distillation}.
\newblock \bibinfo{journal}{\emph{Advances in neural information processing
  systems}}  \bibinfo{volume}{34} (\bibinfo{year}{2021}),
  \bibinfo{pages}{9694--9705}.
\newblock


\bibitem[\protect\citeauthoryear{Liu, Yang, Lei, Wang, Tang, Zhang, Sun, and
  Miao}{Liu et~al\mbox{.}}{2021}]%
        {liu2021pre}
\bibfield{author}{\bibinfo{person}{Yong Liu}, \bibinfo{person}{Susen Yang},
  \bibinfo{person}{Chenyi Lei}, \bibinfo{person}{Guoxin Wang},
  \bibinfo{person}{Haihong Tang}, \bibinfo{person}{Juyong Zhang},
  \bibinfo{person}{Aixin Sun}, {and} \bibinfo{person}{Chunyan Miao}.}
  \bibinfo{year}{2021}\natexlab{}.
\newblock \showarticletitle{Pre-training graph transformer with multimodal side
  information for recommendation}. In \bibinfo{booktitle}{\emph{Proceedings of
  the 29th ACM International Conference on Multimedia}}.
  \bibinfo{pages}{2853--2861}.
\newblock


\bibitem[\protect\citeauthoryear{Radford, Kim, Hallacy, Ramesh, Goh, Agarwal,
  Sastry, Askell, Mishkin, Clark, Krueger, and Sutskever}{Radford
  et~al\mbox{.}}{2021}]%
        {DBLP:journals/corr/abs-2103-00020}
\bibfield{author}{\bibinfo{person}{Alec Radford}, \bibinfo{person}{Jong~Wook
  Kim}, \bibinfo{person}{Chris Hallacy}, \bibinfo{person}{Aditya Ramesh},
  \bibinfo{person}{Gabriel Goh}, \bibinfo{person}{Sandhini Agarwal},
  \bibinfo{person}{Girish Sastry}, \bibinfo{person}{Amanda Askell},
  \bibinfo{person}{Pamela Mishkin}, \bibinfo{person}{Jack Clark},
  \bibinfo{person}{Gretchen Krueger}, {and} \bibinfo{person}{Ilya Sutskever}.}
  \bibinfo{year}{2021}\natexlab{}.
\newblock \showarticletitle{Learning Transferable Visual Models From Natural
  Language Supervision}.
\newblock \bibinfo{journal}{\emph{CoRR}}  \bibinfo{volume}{abs/2103.00020}
  (\bibinfo{year}{2021}).
\newblock
\showeprint[arXiv]{2103.00020}
\urldef\tempurl%
\url{https://arxiv.org/abs/2103.00020}
\showURL{%
\tempurl}


\bibitem[\protect\citeauthoryear{Singh, Hu, Goswami, Couairon, Galuba,
  Rohrbach, and Kiela}{Singh et~al\mbox{.}}{2022}]%
        {singh2022flava}
\bibfield{author}{\bibinfo{person}{Amanpreet Singh}, \bibinfo{person}{Ronghang
  Hu}, \bibinfo{person}{Vedanuj Goswami}, \bibinfo{person}{Guillaume Couairon},
  \bibinfo{person}{Wojciech Galuba}, \bibinfo{person}{Marcus Rohrbach}, {and}
  \bibinfo{person}{Douwe Kiela}.} \bibinfo{year}{2022}\natexlab{}.
\newblock \showarticletitle{Flava: A foundational language and vision alignment
  model}. In \bibinfo{booktitle}{\emph{Proceedings of the IEEE/CVF Conference
  on Computer Vision and Pattern Recognition}}. \bibinfo{pages}{15638--15650}.
\newblock


\bibitem[\protect\citeauthoryear{Wei, Wang, Nie, He, Hong, and Chua}{Wei
  et~al\mbox{.}}{2019}]%
        {wei2019mmgcn}
\bibfield{author}{\bibinfo{person}{Yinwei Wei}, \bibinfo{person}{Xiang Wang},
  \bibinfo{person}{Liqiang Nie}, \bibinfo{person}{Xiangnan He},
  \bibinfo{person}{Richang Hong}, {and} \bibinfo{person}{Tat-Seng Chua}.}
  \bibinfo{year}{2019}\natexlab{}.
\newblock \showarticletitle{MMGCN: Multi-modal graph convolution network for
  personalized recommendation of micro-video}. In
  \bibinfo{booktitle}{\emph{Proceedings of the 27th ACM International
  Conference on Multimedia}}. \bibinfo{pages}{1437--1445}.
\newblock


\end{thebibliography}

\end{document}